# Polarization Bremsstrahlung in collisions of fast ions with multiatomic targets

## M. Ya. Amusia[1, 2] and V. I. Matveev[3]

[1] Racah Institute of Physics, the Hebrew University, Jerusalem 91904, Israel.
[2] Ioffe Physical-Technical Institute, St. Petersburg 194021, Russia.
[3] Lomonosov Nothern (Arctic) federal university, Arkhangelsk 163002, Russia.

**Abstract**
We consider the processes of polarization bremsstrahlung in collisions of fast ions with linear chains consisting of isolated atoms. We obtained intensities and angular distributions of radiation spectra for arbitrary number of atoms in the chain. It appeared that interference in the photon radiation amplitudes lead to prominent variation of spectral angular distributions of polarization bremsstrahlung as compared to these distribuitions in collisions with an isolated atom. The mean loss of energy due to radiation or the so-called rediative friction is estimated.

The results obtained permit standard generalization to the case of polarization bremsstrahlung in fast ion chanelling above surfaces an and in solid body.

## Поляризационное тормозное излучение при столкновениях быстрых ионов с многоатомными мишенями

## М. Я. Амусья[1, 2], В. И. Матвеев[3]

[1] Институт физики им. Дж. Рака, Еврейский университет, Иерусалим 91904, Израиль.
[2] Физико-технический институт им. Иоффе, Санкт-Петербург 194021, Россия.
[3] Северный (Арктический) федеральный университет имени М.В. Ломоносова, 163002 Архангельск, Россия.

**Аннотация**

Рассмотрены процессы поляризационного тормозного излучения при столкновениях быстрых ионов с линейными цепочками, составленными из изолированных атомов. Получены интенсивности и угловые распределения спектров излучения для произвольного числа атомов в цепочке. Показано, что процессы интерференции амплитуд излучения фотонов приводят к заметному изменению спектральных угловых распределений поляризационного тормозного излучения по сравнению с распределениями при столкновениях с изолированным атомом. Результаты допускают стандартное обобщение на случаи поляризационного тормозного излучения при каналировании быстрых ионов над поверхностями и в твердотельных решетках.

**1. Введение**.

В данной работе впервые рассмотрено поляризационное тормозное излучение (ПТИ) не на изолированном атоме, а на многоатомной мишени. ПТИ возникает за счёт того, что налетающий заряд взаимодействует с атомом – мишенью, вызывая поляризацию, в основном дипольную, его электронной оболочки. Наведенный вектор поляризации изменяется в процессе столкновения по величине и направлению, приводя, тем самым, к электромагнитному излучению, которое может превосходить по интенсивности обычное тормозное излучение заряда в поле атома – мишени. Если мишень многоатомна, налетающий ион может поляризовать, в зависимости от угла рассеяния, сразу несколько атомов мишени. В результате, полная амплитуда ПТИ определится суммарным дипольным моментом нескольких атомов мишени. В качестве снаряда мы выбрали ион, поскольку для него ПТИ абсолютно доминирует над обычным тормозным излучением.

ПТИ, сопровождающее столкновения быстрых ионов с отдельными атомами, хорошо изучено [1, 2]. Совершенно иначе обстоит дело с ПТИ, возникающем при столкновении иона с многоатомными мишенями. Например, при выборе в качестве мишеней регулярных многоатомный структур, следует ожидать существенных изменение интенсивности и углового распределения поляризационных фотонов, по сравнению с одноатомными мишенями из-за интерференционных эффектов. Причем, при наличии пространственного разделения со спектрами обычного тормозного излучения можно ожидать значительное увеличения возможностей практических применений эффектов излучения поляризационных фотонов. Отметим, что поляризационное тормозное излучение может быть интерпретировано, как процесс перерассеяния поля иона на атомных электронах, так что проглядывается аналогия с недавно описанными [3] эффектами интерференции при переизлучении ультракоротких импульсов электромагнитного поля линейными многоатомными цепочками.

В настоящей работе рассмотрено поляризационное тормозное излучение, сопровождающее столкновения быстрых ионов с линейными цепочками, составленными из изолированных атомов. Получены интенсивности и угловые распределения спектров излучения для произвольного числа атомов в цепочке. Показано, что процессы интерференции амплитуд излучения приводят к появлению преимущественнного, параллельного линии цепочки, направления вылета фотонов. Другими словами, происходит заметное изменение спектральных интенсивностей и угловых распределений поляризационного тормозного излучения по сравнению с распределениями при столкновениях с изолированным атомом. Результаты допускают обобщение на случаи поляризационного тормозного излучения при рассеянии ионов над многоатомной поверхностью каналировании быстрых ионов в твердотельных решетках.

**2. Вывод формул.**

Рассмотрим сначала потенциал взаимодействия быстрого иона с электроном отдельного атома водорода. Пусть атом помещен в начало системы координат, а ион движется с постоянной нерелятивистской скоростью $v$ и параметром удара $\mathbf{b}$ по траектории $\mathbf{R} = \mathbf{b} + \mathbf{v}t$. Тогда взаимодействие иона с атомным электроном $V(t) = Z/|\mathbf{R}-\mathbf{r}|$ где $\mathbf{r}$ - координаты атомного электрона, $Z$ - заряд иона (здесь и везде ниже используется атомная система единиц). Если атом сдвинут от начала системы координат на расстояние $d$, то его взаимодействие с ионом равно $V(t) = Z/|\mathbf{R}-(\mathbf{r}+\mathbf{d})|$, где $\mathbf{r}$ - координаты атомного электрона, отсчитываемые от ядра атома. Пусть $N$ невзаимодействующих одинаковых одноэлектронных атомов расположены на одной прямой линии на равном расстоянии друг от друга, так что первый атом находится в начале системы координат, а каждый последующий смещен относительно предыдущего на расстояние $d$ вдоль прямой линии. Обозначим как $\mathbf{r}_k$ координаты электрона, принадлежащего атому с номером $k$, координаты $\mathbf{r}_k$ отсчитываются относительно ядра атома с номером $k$, где $k$ принимает значения $k = 1,2,\cdots,N$. Потенциал взаимодействия иона с атомными электронами такой

цепочки атомов равен

$$V = -\sum_{k=1}^{N} \frac{Z}{|\mathbf{R} - (\mathbf{r}_k + (k-1)\mathbf{d})|}. \tag{1}$$

В дипольном приближении, когда $|\mathbf{R} - (k-1)\mathbf{d}| >> |\mathbf{r}_k|$, этот потенциал равен

$$V(t) = \sum_{k=1}^{N} V_k(t), \tag{2}$$

где

$$V_k(t) = \frac{Z}{|\mathbf{R} - (k-1)\mathbf{d}|^3}[\mathbf{R} - (k-1)\mathbf{d}]\mathbf{r}_k. \tag{3}$$

Нам понадобится Фурье-образ этого потенциала

$$\tilde{V}(\omega) = \int_{-\infty}^{+\infty} V(t) e^{i\omega t} dt = \sum_{k=1}^{N} \tilde{V}_k(\omega), \tag{4}$$

где

$$\tilde{V}_k(\omega) = \int_{-\infty}^{+\infty} V_k(t) e^{i\omega t} dt.$$

Пусть $\mathbf{d}$ направлен строго по скорости иона $v$ или, что то же самое, ион движется параллельно цепочке атомов. В этом случае $[\mathbf{R} - (k-1)\mathbf{d}]^2 = [b^2 + v^2(t - (k-1)d/v)]^2$ и легко получить, что

$$\tilde{V}_k(\omega) = -\frac{2Z}{vb}\mathbf{r}_k e^{i\omega(k-1)d/v}\mathbf{F}(\omega),$$

где

$$\mathbf{F}(\omega) = -\frac{\omega b}{v}\left[\frac{\mathbf{v}}{v}K_0\left(\frac{\omega b}{v}\right) - i\frac{\mathbf{b}}{b}K_1\left(\frac{\omega b}{v}\right)\right].$$

Здесь $K_0(x)$ и $K_1(x)$ - модифицированные функции Бесселя. Таким образом Фурье-образ потенциала (2) равен

$$\tilde{V}(\omega) = \sum_{k=1}^{N} \tilde{V}_k(\omega) = -\frac{2Z}{vb}\mathbf{F}(\omega)\sum_{k=1}^{N} \mathbf{r}_k e^{i\omega(k-1)d/v} \tag{5}$$

Нас интересует вероятность излучения одного фотона всеми атомными электронами цепочки при столкновении с движущимся со скорость $v$ ионом заряда $Z$. Амплитуду излучения фотона с импульсом $\mathbf{k}$ и поляризацией $\sigma$ будем вычислять во втором порядке теории возмущений по $V$ и взаимодействию с полем излучения

$$U = -\sum_{k=1}^{N}\left(\frac{2\pi}{\omega}\right)^{1/2}\mathbf{u}_{\kappa\sigma}e^{-i\kappa\mathbf{R}_k}a^{+}_{\kappa\sigma}\hat{\mathbf{p}}_k \;, \qquad (6)$$

где $a^{+}_{\kappa\sigma}$ - оператор рождения фотона с частотой $\omega$, импульсом $\kappa$ и поляризацией $\sigma$, $\mathbf{u}_{\kappa\sigma}$ - единичный вектор поляризации, $\hat{\mathbf{p}}_k = \partial/\partial\mathbf{r}_k$ - операторы импульса атомных электронов, $\mathbf{R}_k = (k-1)\mathbf{d} + \mathbf{r}_k$ - координаты электрона атома $k$ относительно начала системы координат. Координаты $\mathbf{r}_k$ отсчитываются относительно ядра атома с номером $k$, ($k = 1, 2, \cdots, N$).

Во втором порядке теории возмущений по потенциалу $V + U$, переходы с изменением состояний атомных электронов и с излучением фотона возможны лишь для "перекрестных" произведений $VU$ или $VU$.

Ограничимся рассмотрением так называемого упругого поляризационного излучения, в котором вся энергия, теряемая налетающим ионом испускается в виде одного фотона, и везде ниже будем считать, так что все атомы до столкновения и после находятся в основном состоянии. Тогда амплитуда испускания фотона равна

$$c^{(2)} = -i\sum_n \frac{V_{0,n}(\omega)U_{n,0}}{\omega + \omega_{n,0} - i\lambda} + i\sum_n \frac{U_{0,n}V_{n,0}(\omega)}{\omega + \omega_{0,n} + i\lambda}. \qquad (7)$$

Каждый атом считаем находящимся в основном состоянии $\varphi_0(\mathbf{r}_k)$ с энергией $\varepsilon_0$. Тогда волновая функция начального состояния всех $N$ электронов вышеописанной цепочки из $N$ невзаимодействующих одинаковых атомов равна

$$\Phi_0 = e^{-i\varepsilon_0 t}\varphi_0(\mathbf{r}_1)e^{-i\varepsilon_0 t}\varphi_0(\mathbf{r}_2)\cdots e^{-i\varepsilon_0 t}\varphi_0(\mathbf{r}_N), \qquad (8)$$

где координаты $\mathbf{r}_k$ отсчитываются относительно ядра атома с номером $k$, а энергия основного состояния $N\varepsilon_0$. Волновую функцию произвольного возбужденного состояния отдельного атома (имеющего номер $a$ в цепочке) будем обозначать $\varphi_{nk}(\mathbf{r}_k)$. Тогда волновая функция произвольных возбужденных состояний всех $N$ электронов вышеописанной цепочки из $N$ невзаимодействующих одинаковых атомов равна

$$\Phi_n = e^{-i\varepsilon_{n1}t}\varphi_{n1}(\mathbf{r}_1)e^{-i\varepsilon_{n2}t}\varphi_{n2}(\mathbf{r}_2)\cdots e^{-i\varepsilon_{nN}t}\varphi_{nN}(\mathbf{r}_N)\,, \qquad (9)$$

где $n = (n1, n2, \cdots, nN)$ - совокупность квантовых чисел для электронных состояний всех атомов цепочки. Энергия такого возбужденного состояния равна $\varepsilon_{n1} + \varepsilon_{n2} + \ldots + \varepsilon_{nN} - N\varepsilon_0$. Соответственно в (7) $\omega_{n,0} = \varepsilon_{n1} + \varepsilon_{n2} + \ldots + \varepsilon_{nN} - N\varepsilon_0$. В амплитуду (7) входят матричные элементы $V_{n,0} = V_{n1,n2,\ldots,n_N;0,0,\ldots,0}$ и $U_{n,0} = U_{n1,n2,\ldots,n_N;0,0,\ldots,0}$ от операторов, являющихся суммами одночастичных операторов. Матричные элементы от таких операторов отличаются от нуля только для переходов с изменением состояния какого-либо одного электрона. Поэтому, нетрудно убедиться, что амплитуду (7) можно привести к виду

$$c^{(2)} = -\frac{2Z}{vb}\sqrt{2\pi\omega}\mathbf{u}_{\kappa\sigma}\mathbf{F}(\omega)\alpha(\omega)\sum_{k=1}^{N}e^{i\omega(k-1)d/v}\sum_{k'=1}^{N}e^{-i\mathbf{\kappa}(k'-1)\mathbf{d}}\delta_{k,k'}, \qquad (10)$$

где $\alpha(\omega)$ - динамическая поляризуемость основного состояния атома - мишени, а два последних сомножителя представляют собой сумму геометрической прогрессии, так что

$$c^{(2)} = -\frac{2Z}{vb}\sqrt{2\pi\omega}\mathbf{u}_{\kappa\sigma}\mathbf{F}(\omega)\alpha(\omega)\frac{1-e^{iN(\omega d/v-\mathbf{\kappa d})}}{1-e^{i(\omega d/v-\mathbf{\kappa d})}}. \qquad (11)$$

После суммирования $|c^{(2)}|^2$ по поляризациям фотона и интегрирования по углу вектора параметра удара и его длине (от нижнего значения $b_0$ до $\infty$), получаем сечение испускания (всеми атомными электронами цепочки) фотона с импульсом $\mathbf{\kappa}$ в телесный угол $d\Omega_{\mathbf{\kappa}}$, описанный вокруг направления вылета фотона:

$$\frac{d^2\sigma}{d\omega d\Omega_{\mathbf{\kappa}}} = \frac{2}{\pi}\frac{\omega^3}{c^3}\frac{Z^2}{v^2}|\alpha(\omega)|^2\frac{\omega b_0}{2v}\left\{(1-\frac{3}{2}[\hat{\mathbf{v}}\times\hat{\mathbf{k}}]^2)\frac{\omega b_0}{v}K_0^{\,2}\left(\frac{\omega b_0}{v}\right)+\right.$$
$$\left.2(1-\frac{1}{2}[\hat{\mathbf{v}}\times\hat{\mathbf{k}}]^2)K_0\left(\frac{\omega b_0}{v}\right)K_1\left(\frac{\omega b_0}{v}\right) - (1-\frac{3}{2}[\hat{\mathbf{v}}\times\hat{\mathbf{k}}]^2)\frac{\omega b_0}{v}K_1^{\,2}\left(\frac{\omega b_0}{v}\right)\right\}\times \qquad (12)$$
$$\left|\frac{\sin(N\omega d/v - N\mathbf{\kappa d})}{\sin(\omega d/v - \mathbf{\kappa d})}\right|^2,$$

где $\hat{\mathbf{v}}$ и $\hat{\mathbf{k}}$ - единичные вектора, направленные вдоль скорости иона и импульса фотона, соответственно, $[\hat{\mathbf{v}}\times\hat{\mathbf{k}}]$ - их векторное произведение. Нетрудно убедиться, что выражение (12), как это и должно быть для сечения, положительно определено.

Оценка дифференциального сечения испускания тормозного фотона при характерных атомных частотах $\omega\approx 1$ и скоростях $v\ll Z$ имеет вид

$$\frac{d^2\sigma}{d\omega d\Omega_{\mathbf{\kappa}}} = \frac{2}{\pi}\frac{\omega^3}{c^3}\frac{Z^2}{v^2}(1-\frac{1}{2}[\hat{\mathbf{v}}\times\hat{\mathbf{k}}]^2)|\alpha(\omega)|^2\ln\left(\frac{2v}{\omega b_0\gamma}\right)\times\left[\frac{\sin(N\omega d/v - N\mathbf{\kappa d})}{\sin(\omega d/v - \mathbf{\kappa d})}\right]^2, \qquad (13)$$

где $\gamma = 1,781$. Как указывалось при получении формулы (11), величина $b_0$ имеет смысл наименьшего параметра удара до которого справедливо дипольное приближение (3) и ассимптотика при больших значениях аргумента для функций Бесселя, входящих в формулу (12). Другими словами, формула (13) получена с логарифмической точностью, когда не только аргумент у логарифмической функции велик по сравнению с единицей, но и значение логарифма велико. Поэтому, в (13) мы можем положить $b_0\sim 1$ - порядка атомного размера и считать, что под знаком логарифма $2/(b_0\gamma)\approx 1$.

### 3. Обсуждение результатов

Поскольку $|\omega d/v - \kappa d| = \omega d/v(1-v/c)$, в рассматриваемом нами нерелятивистском пределе слагаемым с **kd** в аргументах синусов в (13) можно пренебречь, что приводит в упомянутом выше логарифмическом приближении к формуле

$$\frac{d^2\sigma}{d\omega d\Omega_k} = \frac{2}{\pi}\frac{\omega^3}{c^3}\frac{Z^2}{v^2}(1-\frac{1}{2}[\hat{\mathbf{v}}\hat{\mathbf{k}}]^2)|\alpha(\omega)|^2 \ln\left(\frac{v}{\omega}\right) \times \left[\frac{\sin(N\omega d/v)}{\sin(\omega d/v)}\right]^2. \qquad (14)$$

Таким образом, полученное сечение отличается от сечения ПТИ одного изолированного атома лишь множителем $\eta = [(\sin N\omega d/v)/(\sin \omega d/v)]^2$, который может быть весьма значительным, если $N\omega d/v \cong 1$, а $\omega d/v \ll 1$. В этом случае фактором усиления становится величина $\eta \cong (v/\omega d)^2 \gg 1$, не зависящая от $N$. Если длина цепочки невелика, при большой скорости снаряда возможно выполнение и условия $N\omega d/v \ll 1$, что ведёт к $\eta \approx N^2$.

В действительности число атомов $N$ в цепочке ограничивается требованием применимости использованной нами временной теории возмущений по потенциалу $V$, для чего необходима малость интеграла

$$\int_{-\infty}^{\infty} V(t)dt \approx V\tau \approx (Z/b)Nb/v = NZ/v \ll 1. \qquad (15)$$

Следовательно

$$N \ll v/Z,$$

что накладывает ограничение на максимальное число атомов в цепочке. Очевидно, что это результат использования теории возмущений, поскольку одновременное поляризующее действие налетающего иона на атомы цепочки при малых скоростях $v$ окажется больше, чем при больших.

С ростом $v$ эффекты интерференции начинают проявляться и в угловом распределении, там, где $|\omega d/v - \mathbf{kd}| \to 0$. Для фотонов, вылетающих в направлении оси цепочки, фактор усиления становится большим, т.е. $\eta \approx N^2$. Это проявляется лишь для скоростей налетающего иона, близких к световой, где, на первый взгляд, требуется релятивистский подход. Однако, если рассматривать столкновение движущегося с релятивистской скоростью иона с лёгким нерелятивистским атомом, и оставаться в рамках дипольного приближения, то и после столкновения атомные электроны нерелятивистские. Другими словами, и при релятивистских скоростях, приведенные выше формулы остаются справедливыми.

С помощь (13) или (14) можно оценить полную энергия радиационных потерь на излучеие со стороны налетающего иона, и найти, тем самым, коэффициент радиационного трения. Для этого следует проинтегрировать формулу (13) или (14) по всем телесным углам, умножить на частоту фотона и на число атомов, приходящихся на единицу длины цепоки, после чего проинтегрировать по всем частотам фотона. При интегрировании следует учесть, что верхний предел для частоты $\omega$ есть $\omega < v$. Надо иметь в виду, что в области применимости (14), когда $\omega/v \ll 1$ sin $\sin(\omega d/v) \approx \omega d/v$, поскольку $d \geq 1$.

Используем в качестве поляризуемости амома цепочки её ассимптотическое выражение $\alpha(\omega) \simeq N_C/\omega^2$, где $N_C$ - число электронов в одном атоме цепочки. Считаем это выражение справедливым вплоть до $\omega \geq I_C$, где $I_C$ - потенциал ионизации атома цепочки. При $\omega \leq I_C$ полагаем $\alpha(\omega) \simeq N_C / \omega_C^2$. Обозначая коэффициент трения через $K(v)$, получаем

$$K(v) = \int_0^v \omega \frac{d\sigma}{d\omega} = \frac{Z^2 N_C^2}{c^3 I_C} \ln \frac{v}{I_C}. \qquad (16)$$

При выводе (15) для простоты положено $d = 1$. Оценка радиационной потери на всю длину цепочки получается умножением (16) на $N$. Примечательно, что вклад областей $0 < \omega < I_C$ и $I_C < \omega < v$ оказался по порядку величины одинаков.

### 4. Выводы и заключительные замечания

Мы показали, что интерференционные эффекты при генерации ПТИ при рассеянии ионов на цепочке атомов весьма сильны, приводя к увеличению интесивности излучения, и, с ростом скорости, к его преимущественной ориентации вдоль направления оси цепочки.

Приведенные выше формулы получены в предположении, что налетающий ион не имеет внутренней структуры, и не может поляризоваться вследствие взаимодействия с цепочкой атомов – мишенью. Обощение рассмотренной модели на случай поляризуемого иона или даже нейтрального атома представляет значительный интерес, и может быть достигнуто, следуя методикам, изложенным в [1] и [2].

Весьма интересно излучение, возникающее при движении двух цепочек разных атомов друг относительно друга.

Полученные результаты могут быть обобщены на случай рассеяния иона над многоатомной поверхностью, подобной листу графена, а также рассеяния иона на трёхмерной кристаллической структуре. Определённый интерес представляет и излучение при движении парралельно оси нанотрубки, как снаружи трубки, так и внутри неё.

В целом, простая модель, предложенная в данном письме позволяет получить интересные результаты и допускает перспективные обобщения.

### Литература